\documentstyle[12pt]{article}

\begin{document}
\vspace{-2.0cm}
\bigskip
\begin{center}
{\Large \bf
The  Translation  Groups  as  Generators  of Gauge Transformation in 
B$\wedge$F Theory}
\end{center}
\vskip .8 true cm
\begin{center}
{\bf Rabin Banerjee}\footnote{rabin@boson.bose.res.in} and
{\bf Biswajit Chakraborty}\footnote{biswajit@boson.bose.res.in}

\vskip 1.0 true cm

S. N. Bose National Centre for Basic Sciences \\
JD Block, Sector III, Salt Lake City, Calcutta -700 098, India.
        
\end{center}
\bigskip

\centerline{\large \bf Abstract}
The translation group $T(2)$, contained in Wigner's little group for
massless particles, is shown to generate gauge transformations in the
Kalb-Ramond theory, exactly as happens in Maxwell case. For the 
topologically massive ($B\wedge$F) gauge
theory, both $T(2)$ and $T(3)$,
act as the corresponding generators.

\newpage
Ever since Wigner introduced the concept of little group for massive and massless relativistic particles way back in 1939 \cite{wigner}, it played the most
important role in classifying the various particles according to their spin
quantum number. Furthermore, the transformation properties of quantum
states, belonging to the Hilbert space, under Poincare transformation  
can only be obtained by the method of induced representation from its
corresponding transformation properties under the little group. For example, the little group for a massive particle is $SO(3)$
exactly as its non-relativistic counterpart. This implies that
 the transformation property of a massive particle, under rotation,
 will be the same as that of the 
non-relativistic one allowing only the usual integer or half-integer spin for 
massive particles and the whole apparatus of spherical  harmonics, 
Clebsch-Gordon coefficients etc can be carried over  from
 non-relativistic  to relativistic quantum mechanics\cite{weinberg2}.

The situation for massless particles is quite different on the other hand.
First of all it has no non-relativistic counterpart. Secondly, the structure
 of the corresponding little group E(2),  which is a semi-direct product
of $T(2)$(group of translations in a two dimensional plane perpendicular
to the direction of momentum) and $SO(2)$, 
 is not semi-simple. In fact,
 $T(2) $ is the Abelian 
invariant subgroup of E(2). This
fact gives rise to some interesting complications. One can show that 
the spin has to align itself either in the parallel or anti-parallel direction
of the momentum  of the particle with the allowed values of the helicity,
restricted to 
integers and half integers, just as for its massive counterpart\cite{weinberg2}.
 The question naturally arises 
regarding the role of the other two "translation" like generators of $T(2) \subset$  E(2).
Again it was shown by Weinberg\cite{weinberg,weinberg2} and Han et.al.\cite{kim} that these objects
play the role of generators of gauge transformations in Maxwell theory, which
is a $U(1)$ gauge theory. As is well known, a typical gauge theory like 
Maxwell theory does not allow massive excitations. This is in contrast with 
topologically massive theory like Maxwell-Chern-Simons(MCS) model in 2+1
dimensions\cite{deser}.
 The B$\wedge$F model \cite{CS} is 
another example in the usual 3+1 dimensions  where gauge invariance coexists
with mass.

One can therefore ask whether the translation like generators of $T(2)$
 can also generate gauge transformation in the rest frame of a quanta in
such topologically massive gauge theories? A first 
attempt in this direction was made in \cite{bcs}, where it was found that 
the little group for massless particles in 2+1 dimensions, albeit in a 
different representation, can still generate gauge transformation in the 
MCS theory mentioned earlier. 
But the Wigner's little group for massless particles in 2+1 dimensions
involves only a single parameter(which is isomorphic to ${\cal R} \times {\cal Z}_2$\cite{binegar}), so the construction of the desirable(non-unique)
representation generating the gauge transformation in MCS theory was 
straightforward. The question is what happens in the more complicated 
(3+1) dimensional case? For that we consider Kalb-Ramond(KR) \cite{KR} 
and B$\wedge$F
theory \cite{CS} - both in 3+1 dimension and involving a 2-form gauge field.
As is well known, the KR quanta are massless
unlike the topologically massive B$\wedge$F theory \cite{A}. 

Let us first provide a brief review of the role of the Wigner's little 
group in generating gauge transformation in Maxwell theory.\footnote{In our 
convention, the 4-vectors $x^{\mu} = (t, x, y, z)^T \equiv
(x^0, x^1, x^2, x^3)^T$,  the signature of the metric $g_{\mu \nu} = (+, -, -, -)$ and the 
totally  antisymmetric tensor $\epsilon^{0123}$ = +1.}
The form of the Wigner's little group of a massless particle moving along
$z$-direction is given by \cite{weinberg2,kim},
\begin{equation}
  W(\phi, u, v) = \{ {W^{\mu}}_{\nu} \} = \left( \begin{array}{cccc}
 (1+ \frac{u^2 + v^2}{2}) & (u\cos \phi -v \sin \phi) & (u\sin \phi + v \cos \phi) &
 (-\frac{u^2 + v^2}{2}) \\
u & \cos \phi & \sin \phi & -u \\
v & -\sin \phi & \cos \phi & -v \\
(\frac{u^2 + v^2}{2}) & (u\cos \phi -v \sin \phi) & (u\sin \phi + v \cos \phi) &
(1 - \frac{u^2 + v^2}{2})
\end{array} \right)
\label{eq-1}
\end{equation}

By definition, this preserves the 4-momentum $p^{\mu} = (\omega, 0, 0, \omega)^T$
of the massless particle of energy $\omega$,
\begin{equation}
{W^{\mu}}_{\nu} p^{\nu} = p^{\mu}
\label{eq-2}
\end{equation}
This matrix $ W(\phi, u, v)$ can be factorized as,
\begin{equation}
W(\phi, u, v) = W(0, u, v)R(\phi)
\label{eq-3}
\end{equation}
with,
\begin{equation}
R(\phi) = \left( \begin{array}{cccc}
1 & 0 & 0 & 0 \\
0 & \cos \phi & \sin \phi & 0 \\
0 & -\sin \phi  & \cos \phi & 0 \\
0 & 0 & 0 & 1
\end{array} \right)
\label{eq-4}
\end{equation}
representing the rotation around the $z$-axis - the direction of propagation of
photon and $W(0, u, v)$ is isomorphic to the group of translations $T(2)$ in  2-dimensional Euclidean plane. 
As a whole, the Lie algebra of $W(\phi, u, v)$ is isomorphic to that of
E(2) and ISO(2). 

Now consider a photon having the above mentioned 4-momentum $p^{\mu}$ and
polarization vector $\varepsilon^{\mu}(p)$, so that the negative frequency
 part of the Maxwell gauge field $A^{\mu}(x)$ can be written as,
\begin{equation}
A^{\mu}(x) = \varepsilon^{\mu}(p) e^{ip \cdot x} 
\label{eq-5}
\end{equation}
For the sake of simplicity, we shall suppress the positive frequency part 
and work with (\ref{eq-5}) only. 

Note that a gauge transformation,
\begin{equation}  
A_{\mu}(x) \rightarrow A^{\prime}_{\mu} =  A_{\mu} + \partial_{\mu}f
\label{eq-6}
\end{equation}
for some function $f(x)$ can be written equivalently in  terms of the 
polarization vector $\varepsilon^{\mu}$ as, 
\begin{equation}
\varepsilon_{\mu}(p) \rightarrow \varepsilon^{\prime}_{\mu} =  \varepsilon_{\mu}(p) + if(p)p_{\mu}
\label{eq-7}
\end{equation}
where $f(x)$-a scalar function- has been written as $f(x) = f(p)e^{ip \cdot x}$
just like $A^{\mu}$ (\ref{eq-5}) and again suppressing the positive frequency part. Now the free Maxwell theory has the equation,
\begin{equation}
\partial_{\mu}F^{\mu \nu} = 0
\label{eq-8}
\end{equation}
which follows from the Lagrangian ${\cal{L}} = -\frac{1}{4} F^{\mu \nu}F_{\mu \nu}$
. In terms of the gauge field $A^{\mu}$ the above equation can be rewritten as,
\begin{equation}
\left(g^{\nu}_{\mu} \Box -\partial^{\nu} \partial_{\mu} \right) A^{\mu}=0
\end{equation}
which again can be cast in terms of the polarization vector using (\ref{eq-5}) as,
\begin{equation}
p^2 \varepsilon^{\mu} - p^{\mu} p_{\nu} \varepsilon^{\nu} = 0
\label{eq-9}
\end{equation}
One can easily see at this stage that for $p^2 \ne 0$, the polarization vector $\varepsilon^{\mu}$ is proportional to $p^{\mu}$;
\begin{equation}
\varepsilon^{\mu} = \frac{(p^{\nu}\varepsilon_{\nu})}{p^2}p^{\mu}
\label{eq-10}
\end{equation}
so that using (\ref{eq-7}), one gauges it away by making an appropriate 
choice for $f(p)$,
\begin{equation}
f(p) = \frac{i}{p^2}(p^{\nu}\varepsilon_{\nu})
\end{equation}
One therefore concludes that massive excitations, if any, are gauge artefacts
in pure Maxwell theory. This is not true for massless excitations $p^2 = 0$.
Using (\ref{eq-9}), one finds that this only implies,
\begin{equation}
p_{\mu}\varepsilon^{\mu} = 0
\label{eq-11}
\end{equation}
which is nothing but the Lorentz gauge condition $\partial_{\mu}A^{\mu} = 0$.
So in the frame, where momentum 4-vector takes the form $p^{\mu} = (\omega, 0, 0, \omega)^T$, the corresponding $\varepsilon^{\mu}$ takes the form $\varepsilon^{\mu} = (\varepsilon^0, \varepsilon^1, \varepsilon^2, \varepsilon^0)^T$, which is 
again gauge equivalent to,
\begin{equation}
\varepsilon^{\mu}
 = (0, \varepsilon^1, \varepsilon^2, 0)^T 
\label{eq-12}
\end{equation}
as one can easily show using (\ref{eq-7})\cite{felsager}.   
 The physical polarization vector is thus confined in the
 $xy$ plane for the photon 
moving along the $z$-direction and has only two transverse degrees of freedom.
The other scalar and longitudinal polarization can be gauged away. 

Coming 
finally to the role of $T(2)$ in generating gauge transformations,
one can easily see that the action of the little group element $W(0, u, v)$(\ref{eq-1}) on $\varepsilon^{\mu}$ (\ref{eq-12}) generates the following transformation,
\begin{equation}
\varepsilon^{\mu} \rightarrow \varepsilon^{\prime \mu} = {W^{\mu}}_{\nu}(0, u, v) \varepsilon^{\nu} = \varepsilon^{\mu} +
 \left( \frac{u\varepsilon^1 + v\varepsilon^2}{\omega}\right)p^{\mu}
\label{eq-13}
\end{equation}
Clearly using (\ref{eq-7}), this can be identified as a gauge transformation.
This result was obtained earlier in \cite{weinberg,kim}. 

We next perform a similar analysis for the Kalb-Ramond theory \cite{KR}
whose dynamics is
governed by the Lagrangian, 
\begin{equation}
{\cal{L}} = \frac{1}{12}H_{\mu \nu \lambda}H^{\mu \nu \lambda}
\label{eq-14}
\end{equation}
where
\begin{equation}
H_{\mu \nu \lambda} = \partial_{\mu}B_{\nu \lambda} + \partial_{\nu} B_{\lambda \mu} + \partial_{\lambda} B_{\mu \nu}
\end{equation}
is the rank-3 antisymmetric field strength tensor and is derived from the
rank-2 antisymmetric gauge field 
\begin{equation}
B_{\mu \nu} = -B_{\nu \mu}
\end{equation}
The corresponding equation of motion is given by
\begin{equation}
\partial_\mu H^{\mu \nu \lambda} = 0
\label{eq-16}
\end{equation}
Here the model is invariant under the gauge transformation given by
\begin{equation}
B_{\mu \nu} \rightarrow B^{\prime}_{\mu \nu} = B_{\mu \nu} + \partial_\mu f_\nu - \partial_\nu f_\mu 
\label{eq-15}
\end{equation}
In contrast to the Maxwell theory, these gauge transformations are reducible,
i.e., they are not all independent. This is connected to the fact that 
it is possible to choose some $f_{\mu} = \partial_{\mu}\Lambda$ for which
the gauge variation vanishes trivially.
Proceeding just as was done for the Maxwell case, the negative frequency part for the KR gauge field for a particle of 4-momentum $ p^{\mu}$ can be written as
\begin{equation}
B^{\mu \nu}(x) = \varepsilon^{\mu \nu}(p)e^{i p \cdot x}
\label{eq-17}
\end{equation}
where we have introduced an antisymmetric polarization tensor ($\varepsilon^{\mu \nu} = - \varepsilon^{\nu \mu} $), the counterpart of $\varepsilon^{\mu}$ 
in (\ref{eq-5}). 
One can again cast the gauge transformation (\ref{eq-15}) and the equation of motion (\ref{eq-16}) in terms of the polarization tensors as 
\begin{equation}
\varepsilon_{\mu \nu} \rightarrow \varepsilon^{\prime}_{\mu \nu} = \varepsilon_{\mu \nu} + i(p_{\mu}f_{\nu}(p) - p_{\nu}f_{\mu}(p))
\label{eq-18}
\end{equation}
and 
\begin{equation}
p_{\mu}[p^{\mu} \varepsilon^{\nu \lambda} + p^{\nu} \varepsilon^{\lambda \mu} + p^{\lambda} \varepsilon^{\mu \nu}] =0
\label{eq-19}
\end{equation}
respectively. We can then again consider the massive $(p^2 \ne 0)$ and massless
$(p^2 = 0)$ cases separately. For $(p^2 \ne 0)$, one can write,  
\begin{equation}
\varepsilon^{\nu \lambda} = \frac{1}{p^2}[p^{\nu}(p_{\nu}\varepsilon^{\mu \lambda}) - p^{\lambda}(p_{\mu}\varepsilon^{\mu \nu})] 
\label{eq-20}
\end{equation}
Using (\ref{eq-18}), this can be gauged away by choosing
\begin{equation}
f^{\lambda}(p) = \frac{i}{p^2}p_{\mu}\varepsilon^{\mu \lambda}
\label{eq-21}
\end{equation}
We thus find that massive excitations, if any, are gauge artefacts just as in
the Maxwell case. For $p^2 = 0$, on the other hand, one gets by using (\ref{eq-19})
\begin{equation}
p_{\mu}\varepsilon^{\mu \nu} = 0 
\label{eq-22}
\end{equation}
which is again equivalent to the "Lorentz condition" $\partial_{\mu}B^{\mu \nu} = 0$. Using this condition the six independent components of the 
antisymmetric matrix $\varepsilon \equiv \{\varepsilon^{\mu \nu}\}$ can be 
reduced further. For example, in the frame where the light-like vector $p^\mu$
takes the form $p^\mu = (\omega, 0, 0, \omega)^T$, the condition (\ref{eq-22}) 
reduces to 
\begin{equation}
\varepsilon \cdot p = \left( \begin{array}{cccc}
0 & \varepsilon^{01} & \varepsilon^{02} & \varepsilon^{03} \\
-\varepsilon^{01} & 0 & \varepsilon^{12} & \varepsilon^{13}  \\
-\varepsilon^{02} & -\varepsilon^{12} & 0 & \varepsilon^{23}  \\
-\varepsilon^{03} & -\varepsilon^{13}  & -\varepsilon^{23} & 0
\end{array} \right) \left( \begin{array}{c}
\omega \\
0 \\
0 \\
\omega
\end{array} \right) = 0
\label{eq-23}
\end{equation}
Simplifying, this yields
\begin{equation}
\varepsilon^{03} = 0; \varepsilon^{01} = \varepsilon^{13}; \varepsilon^{02} = \varepsilon^{23} 
\label{eq-24}
\end{equation}
so that the reduced form of the $\varepsilon$ matrix is given by 
\begin{equation}  
\varepsilon = 
\left( \begin{array}{cccc}
0 & \varepsilon^{01} & \varepsilon^{02} & 0 \\
-\varepsilon^{01} & 0 & \varepsilon^{12} & \varepsilon^{01} \\
-\varepsilon^{02} & -\varepsilon^{12} & 0 & \varepsilon^{02} \\
0 & -\varepsilon^{01} & -\varepsilon^{02} & 0
\end{array} \right)
\label{eq-25}
\end{equation}
This form of $\varepsilon$ can be reduced further by making the gauge 
transformation (\ref{eq-18}), by taking,
$f^1 = \frac{i}{\omega} \varepsilon^{01}$ 
and
 $f^2 = \frac{i}{\omega} \varepsilon^{02}$
whereby $\varepsilon^{01}$ and $\varepsilon^{02}$ are gauged  away and one 
is left with just one degree of freedom represented by $\varepsilon^{12}$.
The final form of the $\varepsilon$ matrix is then,
\begin{equation}
\varepsilon = \varepsilon^{12}
  \left( \begin{array}{cccc}
0 & 0 & 0 & 0 \\
0 & 0 & 1 & 0 \\
0 & -1 & 0 & 0 \\
0 & 0 & 0 & 0 
 \end{array} \right)
\label{eq-26}
\end{equation}
We are now in a position to study the role of $T(2)$ in generating 
gauge transformations. Using the fact that the rank-2 contravariant tensor
$\varepsilon^{\mu \nu}$ transforms as 
\begin{equation}
\varepsilon^{\mu \nu} \rightarrow \varepsilon^{\prime \mu \nu} = {\Lambda^{\mu}}_\rho {\Lambda^{\nu}}_\sigma \varepsilon^{\rho \sigma}
\end{equation}
under a Lorentz transformation $\Lambda({\Lambda^{\mu}}_\nu \equiv \frac{\partial x^{\prime \mu}}{\partial x^{\nu}})$, one can write down the transformation
 property of $\varepsilon$ matrix under the little group $W(0, u, v)$
(\ref{eq-1}), which is a subgroup of the Lorentz group, in a matrix form as,
\begin{equation}
\varepsilon \rightarrow \varepsilon^{\prime} = W(0, u, v) \varepsilon W^T (0, u, v) \\
= \varepsilon^{12}\left( \begin{array}{cccc}
0 & -v & u & 0 \\
v & 0 & 1 & v \\
-u & -1 & 0 & -u \\
0 & -v & u & 0 
\end{array} \right) = \varepsilon + \varepsilon^{12}\left( \begin{array}{cccc}
0 & -v & u & 0 \\
v & 0& 0 & v \\
-u & 0 & 0 &  -u \\
0 & -v & u & 0
\end{array} \right)
\end{equation}
This is clearly a gauge transformation, as this can be cast in the 
form of  (\ref{eq-18}) with 
$
f^1 = \frac{iv}{\omega}\varepsilon^{12},   
f^2 = - \frac{iu}{\omega}\varepsilon^{12} $
and $f^3 = f^0$.

We finally take up the case of B$\wedge$F model  which has
 massive excitations induced by the topological term.
The  B$\wedge$F model is described by the Lagrangian \footnote{Note that 
the interaction term differs from the conventional B$\wedge$F term $\epsilon_{\mu \nu \lambda \rho}B^{\mu \nu}F^{\lambda \rho}$ by a four divergence.}\cite{CS},
\begin{equation}
{\cal L} = -\frac{1}{4}F_{\mu \nu}F^{\mu \nu} + \frac{1}{12}H_{\mu \nu \lambda}H^{\mu \nu \lambda} - \frac{m}{6}\epsilon^{\mu \nu \lambda \rho} H_{\mu \nu \lambda} A_{\rho}
\label{eq-27}
\end{equation}
which is obtained by topologically coupling the $B_{\mu \nu}$ field of
 Kalb-Ramond theory(\ref{eq-14}) with the Maxwell field $A_{\mu}$ so that the last term in 
(\ref{eq-27}) does not contribute to the  energy-momentum tensor. The parameter
$m$ in this term is taken to be positive. The Euler-Lagrange  equations for 
$A_{\mu}$ and $B_{\mu \nu}$ fields are given by 
\begin{equation}
\partial_{\mu}F^{\mu \rho} - \frac{m}{6}\epsilon^{\mu \nu \lambda \rho}H_{\mu \nu \lambda} = 0 
\label{eq-28}
\end{equation}
and 
\begin{equation}
\partial_{\mu}H^{\mu \nu \lambda} = \frac{1}{2}m \epsilon^{\rho \sigma \nu \lambda}F_{\rho \sigma}
\label{eq-29}
\end{equation}
respectively. Using the forms (\ref{eq-5}) and (\ref{eq-17}), these coupled 
equations (\ref{eq-28}) and (\ref{eq-29}) can be cast in terms of the 
polarization vectors and polarization tensors as,
\begin{equation}
p^2\varepsilon^{\rho} - p^{\rho}p_{\mu}\varepsilon^{\mu} + 
\frac{i}{2}m \epsilon^{\mu \nu \lambda \rho}p_{\mu}\varepsilon_{\nu \lambda} = 0
\label{eq-30}
\end{equation}
\begin{equation}
p^2\varepsilon^{\mu \nu} + \varepsilon^{\lambda \mu}p_{\lambda}p^{\nu} - 
\varepsilon^{\lambda \nu}p_{\lambda}p^{\mu} + imp_{\rho}\varepsilon_{\sigma}
\epsilon^{\rho \sigma \mu \nu} = 0 
\label{eq-31}
\end{equation}
As was done in the previous section, here too we can consider the massless
($p^2 = 0$) and massive ($p^2 \ne 0$) cases respectively. For massless case
one can easily show using (\ref{eq-30}) that 
\begin{equation}
\epsilon_{\rho \alpha \beta \gamma}p^{\rho}p_{\mu}\varepsilon^{\mu} = 
im(p_\alpha \varepsilon_{\beta \gamma} - p_{\beta}\varepsilon_{\alpha \gamma} +
p_{\gamma}\varepsilon_{\alpha \beta}) 
\label{eq-32}
\end{equation}
Contracting with $p^{\beta}$ on either side yields,
\begin{equation}
p_{\alpha}p^{\beta}\varepsilon_{\beta \gamma} + p_{\gamma}p^{\beta}\varepsilon_{\alpha \beta} = 0 
\label{eq-33}
\end{equation}
Using (\ref{eq-33}) and the masslessness condition ($p^2 = 0$), one can 
immediately see using (\ref{eq-31}) that
\begin{equation}
p_{\rho}\varepsilon_{\alpha}\epsilon^{\rho \alpha \beta \gamma} = 0
\label{eq-34}
\end{equation}
so that any general solution of $\varepsilon_{\alpha}$ can now be written 
as,
 \begin{equation}
\varepsilon_{\alpha} = f(p)p_{\alpha}
\label{eq-35}
\end{equation}
for some function $f(p)$. Using (\ref{eq-7}), one can thus easily see that 
massless excitations, if any, are gauge artefacts now. This is in contrast with the Maxwell and KR models considered earlier, where the massive excitations are 
gauge artefacts. Let us consider the massive case
($p^2 = \theta^2$) now. Going to the rest frame with $p^{\mu} = (\theta, {\bf 0})$, one can relate the spatial components of $\varepsilon^{\mu}$ and
$\varepsilon^{\mu \nu}$ by making use of (\ref{eq-30}) and (\ref{eq-31}) to
get the following coupled equations
 \begin{equation}
\varepsilon^i = -\frac{im}{2\theta}\epsilon^{0ijk}\varepsilon_{jk}
\label{eq-36}
\end{equation}
\begin{equation}
\varepsilon^{ij} =  -\frac{im}{\theta}\epsilon^{0ijk}\varepsilon_{k}
\label{eq-37}
\end{equation}
whereas $\varepsilon^0$ and $\varepsilon^{0i}$ remain arbitrary. However
these can be trivially gauged away by making use of the gauge transformations
(\ref{eq-7}) and (\ref{eq-18}) and the above mentioned form for the four-momentum
$p^{\mu} = (\theta, {\bf 0})$ in the rest frame. On the other hand, the mutual compatibility of the pair of equations (\ref{eq-36}) and (\ref{eq-37}) implies
 that  we must have 
\begin{equation}
\theta = m
\label{eq-38}
\end{equation} 
as we have taken $m > 0$. This indicates that the strength `$m$' of B$\wedge$F
term in (\ref{eq-27}) can be identified as the mass of the quanta in B$\wedge$F
model. With this, (\ref{eq-36}) and (\ref{eq-37}) simplify further and 
one can write $\varepsilon^{\mu \nu}$ and $\varepsilon^{\mu}$ in terms of the 
three independent parameters, 
\begin{equation}
\varepsilon = \{\varepsilon^{\mu \nu}\} = \left( \begin{array}{cccc}
0 & 0 & 0 & 0 \\
0 & 0 & c & -b \\
0 & -c & 0 & a \\
0 & b & -a & 0 
\end{array} \right)
  \label{eq-39}
\end{equation}
and 
\begin{equation}
\varepsilon^{\mu} = -i \left( \begin{array}{c}
0 \\
a \\
b \\
c 
\end{array} \right)
\label{eq-40}
\end{equation}
Before we proceed further  to construct the generators of gauge transformations in B$\wedge$F model, let us try to 
see how the varying number of degrees of freedoms associated with the three
 different models we have considered so far, can be understood through a 
Hamiltonian analysis at one stroke. 

For this it is convenient to consider the B$\wedge$F model (\ref{eq-27}) itself,
as for $m=0$, this model reduces to a system of decoupled Maxwell and KR
models(\ref{eq-14}). Introduce the momenta variables,
\begin{equation}
\pi^{\mu} = \frac{\partial{\cal L}}{\partial {\dot{A}}_{\mu}} = 
F^{\mu 0}
\label{eq-41}
\end{equation}
\begin{equation}
\pi^{\mu \nu} = \frac{\partial{\cal L}}{\partial {\dot{B}}_{\mu \nu}} = 
\frac{1}{2}(H^{0 \mu \nu} - m \epsilon^{0 \rho \mu \nu}A_{\rho})
\label{eq-42}
\end{equation}
conjugate to $A_{\mu}$ and $B_{\mu \nu}$ respectively. Clearly $\pi^0$
and $\pi^{0i}$ vanish 
\begin{equation}
\pi^0 \approx 0; \pi^{0i} \approx 0
\label{eq-43}
\end{equation}
and correspond to the primary constraints of the model. To check 
for any secondary constraints, we have to get hold of the Legendre transformed Hamiltonian, which is given by
\begin{equation}
{\cal H} = \pi^{\mu}{\dot{A}}_{\mu} + \pi^{<\mu \nu>}{\dot{B}}_{< \mu \nu > } + \frac{1}{4}F_{\mu \nu}F^{\mu \nu} - \frac{1}{12}H_{\mu \nu \lambda}H^{\mu \nu \lambda} + \frac{m}{6}\epsilon^{\mu \nu \lambda \rho}
H_{\mu \nu \lambda}A_{\rho}
\end{equation}
where the symbol $< >$ means that the indices within it have to be
 ordered either in an increasing or decreasing order to avoid double
counting. Upon simplification, this takes the final form as,
\begin{equation}
{\cal H} = \frac{1}{2}\left[\pi^i \pi^i + F_{<ij>}F_{<ij>} + (H_{123})^2
\right] + \frac{1}{2}m^2 {\bf A}^2 + m(A_1 \pi_{23} + A_2 \pi_{31} + A_3
\pi_{12}) - A_0 G -B_{0i}G_i
\label{eq-44}
\end{equation}
with
\begin{equation}
G \equiv \partial_i \pi^i + mH_{123} \approx 0
\label{eq-45}
\end{equation}
and 
\begin{equation}
G^i \equiv \partial_j \pi^{ji} \approx 0 
\label{eq-46}
\end{equation}
being the secondary(Gauss) constraints. They are first class and 
generate appropriate gauge transformations.One can easily check that there
are no tertiary constraint in the model. At this stage one can note the 
following points: (i) The first class constraints $G_i$ 
 are reducible as they satisfy 
\begin{equation}
\partial_iG^i = 0 
\label{eq-47}
\end{equation}
so that the number of such independent constraints is only 2. (ii) The
pair $\pi^0$ (\ref{eq-43}), $G$ (\ref{eq-45})(except the $mH_{123}$ term)
 are two first class constraints in the free Maxwell theory and the
 other pair   $\pi^{0i}$ (\ref{eq-43}) and  $G_i$ (\ref{eq-46}), along 
with the reducibility property (\ref{eq-47}), are the first class      
constraints of the KR field. It is thus easy to see the number of degrees of freedom in B$\wedge$F theory can be obtained by just putting together
the number of degrees of freedom in Maxwell and KR theories.
To that end consider free Maxwell theory which has $4 \times 2 = 8$
variables $( A_{\mu}, \pi^{\nu})$ in phase space to begin with.
The two constraints $(\pi^0, G)$ along with two gauge fixing conditions 
will reduce the dimension of the physical phase space to four which
is equivalent to  two degrees of freedom in the configuration space.
Correspondingly the polarization vector takes the form (\ref{eq-12}) 
with only transverse degrees of freedom surviving. 

Taking up the case of KR theory now, it has $6 \times 2 = 12$ variables
in the phase space$(B_{\mu \nu}, \pi^{\mu \nu})$. The total number of 
first class constraints $\pi^{0i}$ (\ref{eq-43}) and $G_i$ (\ref{eq-45})
is 3+2 = 5 now. Along with gauge fixing conditions, the dimension of phase space will reduce to $12 - (5 \times 2) = 2$ so that there is
only one independent variable in the configuration space. Correspondingly
 the polarization matrix $\varepsilon$ involves a single parameter 
(\ref{eq-26}). The number of independent configuration space variables 
in B$\wedge$F theory is therefore just 3. Correspondingly the polarization vector and polarization tensor take the forms (\ref{eq-39}) and (\ref{eq-40}). 

Another way of understanding the degree of freedom count is to recall that
the B$\wedge$F Lagrangian (\ref{eq-27}) can be regarded either as a massive
Maxwell(i.e., Proca) theory or a massive KR theory \cite{ABL, BB}. This can be achieved by 
eliminating once the KR field or, alternatively, the vector field from the 
coupled set of equations (\ref{eq-28}, \ref{eq-29}). Both these theories have
three massive degrees of freedom. It is intriguing to note that this dual
structure appears to be manifested in (\ref{eq-39}) and (\ref{eq-40}), by the
 orthogonality relation, 
 \begin{equation}
\varepsilon^{\mu \nu} \varepsilon_{\nu} = 0
\end{equation}.

Returning to the issue of gauge transformations in the B$\wedge$F theory, 
observe that $W(0, u, v)$ fails to be a generator. Does this mean that
$T(2)$, in contrast to the Maxwell and KR examples, is not a generator 
of gauge transformation in the B$\wedge$F theory? Before discussing this,
consider the matrix, 
\begin{equation}
D(\alpha, \beta, \gamma) = \left( \begin{array}{cccc}
1 & \alpha & \beta & \gamma \\
0 & 1 & 0 & 0 \\
0 & 0 & 1 & 0 \\
0 & 0 & 0 & 1    
\end{array} \right)
\label{eq-48}
\end{equation}
involving three real parameters $\alpha, \beta, \gamma$. This generates 
gauge transformation acting on the polarization vector and polarization tensor
(\ref{eq-39}) and (\ref{eq-40}): 
\begin{equation}
\varepsilon^{\mu} \rightarrow \varepsilon^{\prime \mu} = {D^{\mu}}_{\nu}(\alpha, \beta, \gamma)\varepsilon^{\nu} = \varepsilon^{\mu} - \frac{i}{m}(\alpha a + 
\beta b + \gamma c)p^{\mu}
\label{eq-49}
\end{equation}
\begin{equation}
\varepsilon \rightarrow \varepsilon^{\prime} = D(\alpha, \beta, \gamma) \varepsilon D^T(\alpha, \beta, \gamma) = \varepsilon + 
\left( \begin{array}{cccc}
0 & (\gamma b- \beta c) & (\alpha c - \gamma a) & (\beta a - \alpha b) \\
-(\gamma b- \beta c) & 0 & 0 & 0 \\
-(\alpha c - \gamma a) & 0 & 0 & 0 \\
-(\beta a - \alpha b) & 0 & 0 & 0 
\end{array} \right)
\label{eq-50}
\end{equation}
as both (\ref{eq-49}) and (\ref{eq-50}) can be easily cast into the form  (\ref{eq-7}) and (\ref{eq-18}) with
appropriate choices of $f(p)$ and $f^i(p)$. Also it preserves the 4-momentum
of a massive particle at rest.
We can now identify the group to which $D(\alpha, \beta, \gamma)$ belongs.
One can easily show that 
\begin{equation}
D(\alpha, \beta, \gamma) \cdot D(\alpha^{\prime}, \beta^{\prime}, \gamma^{\prime}) = D(\alpha + \alpha^{\prime}, \beta + \beta^{\prime}, \gamma + \gamma^{\prime})
\label{eq-51}
\end{equation}
and
\begin{equation}
[P_1, P_2] = [P_1, P_3] = [P_2, P_3] = 0 
\label{eq-52}
\end{equation}
where
\begin{equation}
P_1 = \frac{\partial D(\alpha, 0, 0)}{\partial \alpha}; P_2= \frac{\partial D(0, \beta, 0)}{\partial \beta}; P_3 = \frac{\partial D(0, 0, \gamma)}{\partial \gamma}
\end{equation}
can be thought of as three mutually commuting "translational" generators.
The group can therefore be identified with $T(3)$ - the invariant subgroup
of E(3)\cite{wkt} or ISO(3). 

Clearly three different canonical 
embeddings of $T(2)$ within $T(3)$ can be obtained by
successively setting  one of the parameters $\alpha, \beta, \gamma$ 
to be zero in 
(\ref{eq-48}). Not only that these different $T(2)$'s generate gauge 
transformations, as can be seen trivially from (\ref{eq-49}) and (\ref{eq-50}),
they also preserve the 4-momenta of massless particles moving in $x$, $y$ and
$z$ directions respectively. The only distinguishing features of these
different $T(2)$'s are their representations, which they inherit either from the
Wigner's little group in (\ref{eq-1}) or from the representation of $T(3)$
in (\ref{eq-48}). In the former case, it acts as a generator in Maxwell or
KR theory involving only massless excitations while in the latter case,
it acts as a generator in topologically massive B$\wedge$F theory.
The group $T(3)$, on the other hand, acts as a generator only in the
B$\wedge$F theory.

To conclude, we have found that in B$\wedge$F theory the generator of gauge
transformations can be identified to $T(3)$ - the group of translations in  
$R^3$. Clearly $T(2) \subset T(3)$ also does the job and since it is 
isomorphic to the group defined by $W(0, u, v)$ (\ref{eq-1})  one can say that
the abelian invariant subgroup $T(2)$ of Wigner's little group acts also
as a generator in massive B$\wedge$F theory.
The B$\wedge$F theory therefore manifests both aspects of gauge invariance,
conventional or massive. This does not happen in the usual gauge theories like Maxwell
or KR, where only Wigner's little group T(2) acts as the generator.

Our results may be compared with a dynamical (hamiltonian) approach where
the Gauss operator generates the gauge transformations. The structure of
this operator differs from theory to  theory. On the contrary, our analysis
revealed a universal structure for the generators; namely, the translational group. We feel this is related to the fact that the abelian gauge transformations
are translational in nature. It would be worthwhile to extend this analysis
where the gauge transformations are curvilinear, as in nonabelian theories
or gravity \cite{ng}.

\end{document}